\documentclass[a4paper,12pt]{article}
\usepackage{amsmath,amssymb,amsthm,epic,eepic}
\usepackage[english]{babel}
\title{Integrable non-commutative equations on quad-graphs.
The consistency approach}
\author{ A.I.\,Bobenko
 \thanks{Institut f\"ur Mathematik, Technische Universit\"at Berlin,
 Str. des 17. Juni 136, 10623 Berlin, Germany. E--Mail: {\tt
 bobenko}@{\tt math.tu-berlin.de}}
\and Yu.B.\,Suris
 \thanks{Institut f\"ur Mathematik, Technische Universit\"at Berlin,
 Str. des 17. Juni 136, 10623 Berlin, Germany. E--Mail: {\tt
 suris}@{\tt sfb288.math.tu-berlin.de}}}
\date{}

\def\a{\alpha}
\def\b{\beta}

\def\({\left(}
\def\){\right)}
\def\<{\langle}
\def\>{\rangle}
\def\wx{\widetilde{x}}

\newtheorem{theorem}{Theorem}

\newcommand{\zz}{{\mathfrak z}}
\newcommand{\cA}{{\cal A}}
\newcommand{\cC}{{\cal C}}
\newcommand{\cD}{{\cal D}}
\newcommand{\cE}{{\cal E}}
\newcommand{\cK}{{\cal K}}
\newcommand{\cL}{{\cal L}}
\newcommand{\cM}{{\cal M}}

\newcount\xx   \newcount\xy    \newcount\XX   \newcount\XY
\newcount\yx   \newcount\yy    \newcount\YX   \newcount\YY
\newcount\zx   \newcount\zy    \newcount\ZX   \newcount\ZY
\newcount\xyx  \newcount\xyy   \newcount\XYX  \newcount\XYY
\newcount\xzx  \newcount\xzy   \newcount\XZX  \newcount\XZY
\newcount\yzx  \newcount\yzy   \newcount\YZX  \newcount\YZY
\newcount\xyzx \newcount\xyzy  \newcount\XYZX \newcount\XYZY
\newcount\MM \newcount\MD
\newcount\ox \newcount\oy
\def\summ(#1,#2,#3){#1=#2 \advance#1 #3}
\def\calcshift(#1,#2)
 {#1=#2 \multiply#1 by \MM \divide#1 by \MD \advance#1 -#2 \divide#1 2}
\def\calc(#1,#2,#3)
 {#1=#2 \multiply#1 by \MM \divide#1 by \MD \advance#1 -#3}

\begin{document}
\maketitle
\begin{abstract} We extend integrable systems on quad-graphs, such as the
Hirota equation and the cross-ratio equation, to the non-commutative
context, when the fields take values in an arbitrary associative algebra.
We demonstrate that the three-dimensional consistency property 
remains valid in this case. We derive the non-commutative zero curvature
representations for these systems, based on the latter property. Quantum
systems with their quantum zero curvature representations are particular
cases of the general non-commutative ones.
\end{abstract}

\section{Introduction}\label{sect:intro}

The idea to use the $(d+1)$--dimensional consistency (or 
compatibility) of the discrete $d$--dimensional equations as the definition 
of their integrability was recently put forward in \cite{BS}, \cite{ABS} (and 
independently in \cite{N}). This definition, apart of being 
conceptually transparent, has also other important theoretical advantages. 
So, finding the zero curvature representation for a given discrete 
system becomes an algorithmically solvable problem (recall that normally this 
was considered as a transcendental task whose successful solution is only 
possible with a large portion of luckyness in the guesswork). Also, in 
\cite{ABS} it was 
demonstrated that the consistency criterium can be successfully used to 
classify integrable systems within certain ans\"atze. 

In the present paper we give a further application of the consistency
approach: we show that it works equally smoothly for {\it non-commutative}
equations, where the participating fields live in an arbitrary associative
(not necessary commutative) algebra $\cA$ (over the field $\cK$), 
and not just in ${\mathbb C}$, as in \cite{BS}, 
\cite{ABS}. We do not develop the corresponding classification, but rather 
consider several important examples, which generalize those already appeared 
in various applications, such as the
quantum Hirota equation or the quaternionic cross--ratio equation. It turns
out that finding the zero curvature representation in $2\times 2$ matrices 
with entries from $\cA$ does not hinge on the particular algebra $\cA$ or on 
prescribing some particular commutation rules for fields in the neighboring
vertices (like Weyl commutation relations in the traditional treatment of the
quantum Hirota equation \cite{FV}). The fact that some commutation relations 
are preserved by the evolution, is thus conceptually separated from the
integrability.

\section{Basic setup}\label{sect:basics}

We start with a planar quad--graph $\cD$, i.e. a cell decomposition
of a surface, with all 2--cells being quadrilaterals. The sets of the vertices,
edges, and faces of $\cD$ (i.e. of its 0-, 1-, and 2-cells) will be denoted 
by $V(\cD)$, $E(\cD)$, and $F(\cD)$, respectively. The quad--graph $\cD$ is
supposed to carry a {\it labelling}, i.e. a function $\a$ on its edges which
takes equal values on any two opposite edges of any elementary quadrilateral.
The fields $x\in\cA$ are assigned to vertices of $\cD$. They take values in
an arbitrary associative (in general non-commutative) algebra $\cA$ with unit
over the field $\cK$.

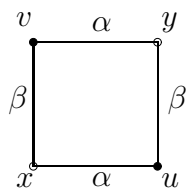
\begin{figure}[htbp]
\begin{center}
\setlength{\unitlength}{0.04em}
\begin{picture}(200,140)(-50,-20)
  \put(100,  0){\circle*{6}} \put(0  ,100){\circle*{6}}
  \put(  0,  0){\circle{6}}  \put(100,100){\circle{6}}
  \put( 0,  0){\line(1,0){100}}
  \put( 0,100){\line(1,0){100}}
  \put(  0, 0){\line(0,1){100}}
  \put(100, 0){\line(0,1){100}}
  \put(-14,-17){$x$}
  \put(103,-17){$u$}
  \put(103,113){$y$}
  \put(-14,113){$v$}
  \put(47,-16){$\a$}
  \put(47,109){$\a$}
  \put(-20,47){$\b$}
  \put(108,47){$\b$}
\end{picture}
\caption{An elementary quadrilateral}\label{Fig:quadrilateral}
\end{center}
\end{figure}

Basic building blocks of systems on quad--graphs are equations on
quadrilaterals of the type
\begin{equation}\label{basic eq}
Q(x,u,v,y;\alpha,\beta)=0,
\end{equation}
where $x,u,v,y\in\cA$ are the fields assigned to the four
vertices of the quadrilateral, and $\alpha,\beta\in\cK$ are 
the parameters assigned to its edges, as shown on
Fig.\,\ref{Fig:quadrilateral}.
We say that the equation (\ref{basic eq}) admits a {\it zero curvature 
representation} if to every oriented edge $(x,u)$ carrying the label $\a$ 
there corresponds a matrix $L(u,x,\a,\lambda)$ depending on an arbitray 
(spectral) parameter $\lambda\in\cK$ such that 
\begin{equation}\label{zero curv cond inverse}
L(x,u,\a,\lambda)=(L(u,x,\a,\lambda))^{-1},
\end{equation}
and for any elementary quadrilateral, as on Fig. \ref{Fig:quadrilateral}, 
the equation (\ref{basic eq}) is equivalent to
\begin{equation}\label{zero curv cond}
L(x,v,\b,\lambda)L(v,y,\a,\lambda)L(y,u,\b,\lambda)L(u,x,\a,\lambda)=I,
\end{equation}
that is,
\begin{equation}\label{zero curv}
L(y,u,\b,\lambda)L(u,x,\a,\lambda)=L(y,v,\a,\lambda)L(v,x,\b,\lambda).
\end{equation}

\section{Non-commutative Hirota equation}
\label{sect:Hirota}

We start our considerations with the following equation:
\begin{equation}\label{Hirota type eq}
 yx^{-1}=f_{\a\b}(uv^{-1}).
\end{equation}
We require that this equation does not depend on how we regard the elementary
quadrilateral. First of all, in general the elementary quadrilaterals are
not supposed to be oriented in some consistent manner, which means that we
cannot distinguish between left and right, so that (\ref{Hirota type eq}) 
has to be equivalent to
\[
 yx^{-1}=f_{\b\a}(vu^{-1}).
\]
Therefore, we require that
\begin{equation}\label{Hirota f cond 1}
f_{\a\b}(A)=f_{\b\a}(A^{-1}).
\end{equation}
Second, the equation (\ref{Hirota type eq}) should allow to exchange the
roles of $x$ and $y$, i.e. to be equivalent to
\[
 xy^{-1}=f_{\a\b}(vu^{-1}).
\]
Hence, we impose the following condition on the function $f_{\a\b}$:
\begin{equation}\label{Hirota f cond 2}
f_{\a\b}(A^{-1})=(f_{\a\b}(A))^{-1}.
\end{equation}
Additionally, if one wants be able to exchange the roles of the pairs $(x,y)$
and $(u,v)$, then (\ref{Hirota type eq}) should be equivalent to
\[
 uv^{-1}=f_{\b\a}(xy^{-1}).
\]
This leads to the following condition on the function $f_{\a\b}$:
\begin{equation}\label{Hirota f cond 3}
f_{\b\a}(A)=f^{-1}_{\a\b}(A^{-1}).
\end{equation}
Here $f_{\a\b}^{-1}$ stands for the inverse function to $f_{\a\b}$, which has
to be distinguished from the inversion in the algebra $\cA$ in the formula 
(\ref{Hirota f cond 2}). 

All the conditions (\ref{Hirota f cond 1})--(\ref{Hirota f cond 3}) are
satisfied for the function which characterizes the {\it Hirota equation}:
\begin{equation}\label{Hirota f}
f_{\a\b}(A)=\frac{1-(\b/\a)A}{(\b/\a)-A}.
\end{equation}

\subsection{Three--dimensional consistency}

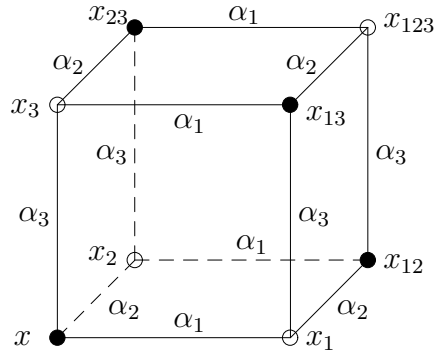
\begin{figure}[htbp]
\begin{center}
\setlength{\unitlength}{0.05em}
\begin{picture}(200,220)(0,0)
 \put(0,0){\circle*{10}}    \put(150,0){\circle{10}}
 \put(0,150){\circle{10}}   \put(150,150){\circle*{10}}
 \put(50,200){\circle*{10}} \put(200,200){\circle{10}}
 \put(50,50){\circle{10}}   \put(200,50){\circle*{10}}
 \path(0,0)(150,0)       \path(0,0)(0,150)
 \path(150,0)(150,150)   \path(0,150)(150,150)
 \path(0,150)(50,200)    \path(150,150)(200,200)   \path(50,200)(200,200)
 \path(200,200)(200,50) \path(200,50)(150,0) 
 \dashline[+30]{10}(0,0)(50,50)
 \dashline[+30]{10}(50,50)(50,200)
 \dashline[+30]{10}(50,50)(200,50)
 \put(-28,-5){$x$} \put(-30,145){$x_3$}
 \put(160,-5){$x_1$} \put(160,140){$x_{13}$}
 \put(210,45){$x_{12}$} \put(210,200){$x_{123}$}
 \put(20,50){$x_2$}  \put(20,205){$x_{23}$}
 \put(75,7){$\a_1$} \put(33,15){$\a_2$}
 \put(115,57){$\a_1$} \put(180,15){$\a_2$}
 \put(-3,173){$\a_2$} \put(75,135){$\a_1$}
 \put(147,173){$\a_2$} \put(110,205){$\a_1$}
 \put(-25,75){$\a_3$}\put(155,75){$\a_3$}
 \put(25,115){$\a_3$}\put(205,115){$\a_3$}
\end{picture}
\caption{Elementary cube of the three--dimensional lattice}\label{cube}
\end{center}
\end{figure}

Now we demonstrate that the non--commutative Hirota equation has a deep 
property of the three--dimensional consistency \cite{BS}.
Consider an elementary cube of the three--dimensional lattice, as shown on 
Fig.\,\ref{cube}.

We assume that all edges of the elementary cube parallel to the axis number
$j$ ($j=1,2,3)$ carry the label $\a_j$. Now, the fundamental 
{\it three--dimensional consistency} property should be 
understood as follows. Suppose that the values of the dependent variable are 
given at the vertex $x$ and at its three neighbors $x_1$, $x_2$, and $x_3$.
Then the Hirota equation (\ref{Hirota type eq}) uniquely determines its
values at $x_{12}$, $x_{13}$, and $x_{23}$. After that the Hirota equation 
delivers three {\it \'a priori} different values for the value of the 
dependent variable at the vertex $x_{123}$, coming from the faces
$(x_1,x_{12},x_{123},x_{13})$, $(x_2,x_{23},x_{123},x_{12})$, and 
$(x_3,x_{13},x_{123},x_{23})$, respectively. The three--dimensional 
consistency means that {\it these three values for $x_{123}$ actually 
coincide}.

\begin{theorem}\label{Hirota 3-dim compatibility} 
The non--commutative Hirota equation is three--dimensionally consistent.
\end{theorem}
\noindent
{\bf First proof of Theorem \ref{Hirota 3-dim compatibility}.} 
We give two proofs of this theorem. The first one is based on direct 
computations and is therefore more specific for the Hirota equation. 
The second one paves the road to the derivation of the zero
curvature representation from the equations governing the system, and is
of a more general nature.

We have, by construction,
\[
x_{ij}x^{-1}=f_{\a_i\a_j}(x_ix_j^{-1}),
\]
and
\begin{equation}\label{Hirota x123}
x_{123}x_i^{-1}=f_{\a_j\a_k}(x_{ij}x_{ik}^{-1}),
\end{equation}
where $(i,j,k)$ is an arbitrary permutation of $(1,2,3)$. So, the
three--dimensional consistency is equivalent to the equation
\[
f_{\a_j\a_k}(x_{ij}x_{ik}^{-1})x_i=f_{\a_i\a_k}(x_{ij}x_{jk}^{-1})x_j,
\]
or else to
\begin{eqnarray*}
\lefteqn{
f_{\a_j\a_k}\Big(f_{\a_i\a_j}(x_ix_j^{-1})(f_{\a_i\a_k}(x_ix_k^{-1}))^{-1}\Big)
=}\\
&&
f_{\a_i\a_k}\Big(f_{\a_i\a_j}(x_ix_j^{-1})(f_{\a_j\a_k}(x_jx_k^{-1}))^{-1}\Big)
x_jx_i^{-1}.
\end{eqnarray*}
Taking into account that actually $f_{\a\b}$ dependes only on 
$\b/\a$, we slightly abuse the notations and write $f_{\a\b}=f_{\b/\a}$.
Denoting $\lambda=\a_j/\a_i$, $\mu=\a_k/\a_j$, and $A=x_ix_j^{-1}$,
$B^{-1}=x_jx_k^{-1}$, and taking into account the property 
(\ref{Hirota f cond 2}), we rewrite the above equation as
\begin{equation}\label{Hirota to prove}
f_{\mu}\Big(f_{\lambda}(A)f_{\lambda\mu}(BA^{-1})\Big)=
f_{\lambda\mu}\Big(f_{\lambda}(A)f_{\mu}(B)\Big)A^{-1}.
\end{equation}
So, in order to prove the theorem, we have to demonstrate that the function
(\ref{Hirota f}) satisfies this functional equation for any 
$\lambda,\mu\in\cK$ and for any $A,B\in\cA$. In this proof we 
repeatedly use the identity
\[
f_{\lambda}(CD^{-1})=(D-\lambda C)(\lambda D-C)^{-1}.
\]
The proof of the functional equation (\ref{Hirota to prove}) is as follows.
\begin{eqnarray*}
\lefteqn{f_{\mu}\Big(f_{\lambda}(A)f_{\lambda\mu}(BA^{-1})\Big)}\\
& = & \Big((\lambda\mu A-B)-\mu f_{\lambda}(A)(A-\lambda\mu B)\Big)
\Big(\mu(\lambda\mu A-B)- f_{\lambda}(A)(A-\lambda\mu B)\Big)^{-1}\\
& = & \Big(\mu A(\lambda-f_{\lambda}(A))-(1-\lambda\mu^2 f_{\lambda}(A))B\Big)
\Big(A(\lambda\mu^2-f_{\lambda}(A))-\mu(1-\lambda f_{\lambda}(A))B\Big)^{-1}
\end{eqnarray*}
Next, we use the fact that
\[
1-\lambda f_{\lambda}(A)=A(\lambda-f_{\lambda}(A)).
\]
This allows us to continue the chain of equations above:
\begin{eqnarray*}
& = & \Big(\mu(1-\lambda f_{\lambda}(A))-(1-\lambda\mu^2 f_{\lambda}(A))B\Big)
\Big(A(\lambda\mu^2-f_{\lambda}(A))-\mu A(\lambda-f_{\lambda}(A))B\Big)^{-1}\\
& = & \Big(\mu-B-\lambda\mu f_{\lambda}(A)(1-\mu B)\Big)
\Big(\lambda\mu(\mu-B)-f_{\lambda}(A)(1-\mu B)\Big)^{-1}A^{-1}\\
& = & f_{\lambda\mu}\Big(f_{\lambda}(A)f_{\mu}(B)\Big)A^{-1}.
\end{eqnarray*}
Theorem is proved. \qed
\medskip

{\bf Remark.} It is difficult to write down an expression for 
$x_{123}$ from which the symmetry with respect to permutations of indices 
$(1,2,3)$ would be apparent. For example, by simplifying (\ref{Hirota x123})
one can get:
\begin{eqnarray}
x_{123} & = & \Big(\frac{\a_j}{\a_i}-x_ix_j^{-1}\Big)^{-1}\Big(
\ell_{ij}x_i+\ell_{jk}x_k+\ell_{ki}x_ix_j^{-1}x_k\Big)\nonumber\\
& \times & \Big(
\ell_{kj}x_i+\ell_{ik}x_j+
\ell_{ji}x_k\Big)^{-1}\Big(\frac{\a_j}{\a_i}-x_ix_j^{-1}\Big)x_j\,,
\end{eqnarray}
where
\[
\ell_{ij}=\frac{\a_i}{\a_j}-\frac{\a_j}{\a_i}.
\]
Of course, in the commutative case this expression becomes symmetric:
\[
x_{123}=\frac{\ell_{ij}x_ix_j+\ell_{jk}x_jx_k+\ell_{ki}x_kx_i}
{\ell_{kj}x_i+\ell_{ik}x_j+\ell_{ji}x_k},
\]
but it is not so in the non--commutative case. On the other hand, one can
rewrite (\ref{Hirota x123}) as
\[
x_{ij}x_{ik}^{-1}=f_{\a_j\a_k}(x_{123}x_i^{-1}),
\]
and a product of three such equations gives:
\[
f_{\a_1\a_2}(x_{123}x_3^{-1})f_{\a_3\a_1}(x_{123}x_2^{-1})
f_{\a_2\a_3}(x_{123}x_1^{-1})=1.
\]
This equation for $x_{123}$ is obviously symmetric with respect to permutations
of indices. Moreover, it makes apparent that $x_{123}$ depends on
$x_1,x_2,x_3$ only, and not on $x$ (this was called the ``tetrahedron
property'' in \cite{ABS}).

\subsection{Zero curvature representation \\
from three--dimensional consistency}
{\bf Second proof of Theorem \ref{Hirota 3-dim compatibility}.}
We have to prove that the following three schemes for
computing $x_{123}$ lead to one and the same result:
\begin{itemize}
\item $(x,x_1,x_2)\mapsto x_{12}\,,$ $\;(x,x_1,x_3)\mapsto x_{13}\,,$
      $\;(x_1,x_{12},x_{13})\mapsto x_{123}\,.$
\item $(x,x_1,x_2)\mapsto x_{12}\,,$ $\;(x,x_2,x_3)\mapsto x_{23}\,,$
      $\;(x_2,x_{12},x_{23})\mapsto x_{123}\,.$ 
\item $(x,x_1,x_3)\mapsto x_{13}\,,$ $\;(x,x_2,x_3)\mapsto x_{23}\,,$
      $\;(x_3,x_{13},x_{23})\mapsto x_{123}\,.$    
\end{itemize}
We shall do this for the first two schemes only, since the rest is done
similarly (or just by changing indices). The Hirota equation on 
the face $(x,x_1,x_{13},x_3)$,
\[
x_{13}x^{-1}=f_{\a_3\a_1}(x_3x_1^{-1}),
\]
can be written as a formula which gives $x_{13}$ as a fractional--linear
transformation of $x_3$:
\begin{equation}\label{Hirota proof aux13}
x_{13}=(\a_1 x_3-\a_3 x_1)(\a_3 x_3-\a_1 x_1)^{-1}x=
L(x_1,x,\a_1,\a_3)[x_3],
\end{equation}
where
\begin{equation}
L(x_1,x,\a_1,\a_3)=\left(\begin{array}{cc}
\a_1 & -\a_3 x_1 \\ \a_3 x^{-1} & -\a_1 x^{-1}x_1\end{array}\right).
\end{equation}
We use here the notation which is common for M\"obius transformations
on ${\mathbb C}$ represented as a linear action of the group 
${\rm GL}(2,\mathbb C)$. In the present case we define the action of the
group ${\rm GL}(2,\cA)$ on $\cA$ by the formula
\[
\left(\begin{array}{cc} a & b \\ c & d \end{array}\right)[z]=
(az+b)(cz+d)^{-1},\qquad a,b,c,d,z\in\cA.
\]
It is easy to see that this is indeed the left action of the group, provided
the multiplication in ${\rm GL}(2,\cA)$ is defined by the natural formula
\[
\left(\begin{array}{cc} a' & b' \\ c' & d' \end{array}\right)
\left(\begin{array}{cc} a & b \\ c & d \end{array}\right)=
\left(\begin{array}{cc} a'a+b'c & a'b+b'd \\ c'a+d'c & c'b+d'd 
\end{array}\right).
\]
Absolutely similarly to (\ref{Hirota proof aux13}), we find:
\begin{equation}\label{Hirota proof aux23}
x_{23}=L(x_2,x,\a_2,\a_3)[x_3].
\end{equation}
From (\ref{Hirota proof aux23}) we derive, by the shift in the direction
of the first coordinate axis, the expression for $x_{123}$ obtained by
the first scheme above:
\begin{equation}\label{Hirota proof aux123'}
x_{123}=L(x_{12},x_1,\a_2,\a_3)[x_{13}],
\end{equation}
while from (\ref{Hirota proof aux13}) we find the expression for $x_{123}$
corresponding to the second scheme:
\begin{equation}\label{Hirota proof aux123''}
x_{123}=L(x_{12},x_2,\a_1,\a_3)[x_{23}].
\end{equation}
Substituting (\ref{Hirota proof aux13}), (\ref{Hirota proof aux23}) on the
right--hand sides of (\ref{Hirota proof aux123'}), 
(\ref{Hirota proof aux123''}), respectively, we represent the equality 
we want to demonstrate in the following form:
\begin{eqnarray}\label{Hirota proof to prove}
\lefteqn{L(x_{12},x_1,\a_2,\a_3)L(x_1,x,\a_1,\a_3)[x_3]}\nonumber\\
& = & L(x_{12},x_2,\a_1,\a_3)L(x_2,x,\a_2,\a_3)[x_3].
\end{eqnarray}
We demonstrate that actually the stronger claim holds, namely that
\begin{equation}\label{Hirota proof to prove 1}
L(x_{12},x_1,\a_2,\a_3)L(x_1,x,\a_1,\a_3)=L(x_{12},x_2,\a_1,\a_3)
L(x_2,x,\a_2,\a_3).
\end{equation}
Indeed, the 11 entries on both parts of this matrix identity are equal
to $\a_1\a_2-\a_3^2x_{12}x^{-1}$. Equating 12 entries on both parts is
equivalent to the Hirota equation of the face $(x,x_1,x_{12},x_2)$, and the
same holds for the 21 entries. Finally, equating the 22 entries is 
equivalent to the condition that $x_{12}x^{-1}$ commutes with $x_2x_1^{-1}$,
and this is, of course, so in virtue of the Hirota equation. This finishes
the second proof. \qed\medskip

Actually, Eq. (\ref{Hirota proof to prove 1}) is nothing but the zero
curvature representation of the non--commutative Hirota equation. It
remains only to spell out the necessary construction which parallels the
commutative one presented in \cite{BS}.

To derive a zero--curvature representation for an equation on $\cD$ of the
type (\ref{basic eq}) possessing the property of the three--dimensional
consistency, we extend the quad--graph $\cD$ into the third dimension. 
This means that we consider the second copy $\cD'$ of $\cD$ and add edges 
connecting each vertex $x\in V(\cD)$ with its copy $x'\in V(\cD')$. 
On this way we obtain a ``three--dimensional quad--graph'' 
${\bf D}$, whose set of vertices is 
\[
V({\bf D})=V(\cD)\cup V(\cD'),
\] 
whose set of edges is 
\[
E({\bf D})=E(\cD)\cup E(\cD')\cup\{(x,x'):x\in V(\cD)\},
\] 
and whose set of faces is 
\[
F({\bf D})=F(\cD)\cup F(\cD')\cup\{(x,u,u',x'):x,u\in V(\cD)\}.
\]
We extend the labelling to $E(\bf D)$ in the following way: each edge
$(x',u')\in E(\cD')$ carries the same label as its counterpart
$(x,u)\in E(\cD)$, while all ``vertical'' edges $(x,x')$ carry one and the
same label $\lambda$. This label plays the role of the spectral parameter. 

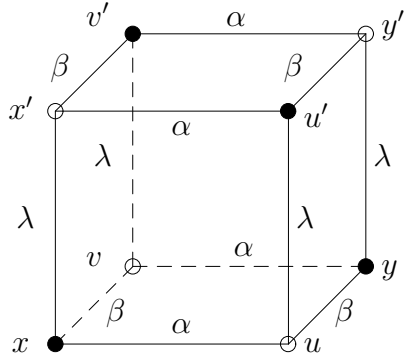
\begin{figure}[htbp]
\begin{center}
\setlength{\unitlength}{0.05em}
\begin{picture}(200,220)(0,0)
 \put(0,0){\circle*{10}}    \put(150,0){\circle{10}}
 \put(0,150){\circle{10}}   \put(150,150){\circle*{10}}
 \put(50,200){\circle*{10}} \put(200,200){\circle{10}}
 \put(50,50){\circle{10}}   \put(200,50){\circle*{10}}
 \path(0,0)(150,0)       \path(0,0)(0,150)
 \path(150,0)(150,150)   \path(0,150)(150,150)
 \path(0,150)(50,200)    \path(150,150)(200,200)   \path(50,200)(200,200)
 \path(200,200)(200,50) \path(200,50)(150,0) 
 \dashline[+30]{10}(0,0)(50,50)
 \dashline[+30]{10}(50,50)(50,200)
 \dashline[+30]{10}(50,50)(200,50)
 \put(-28,-5){$x$} \put(-30,145){$x'$}
 \put(160,-5){$u$} \put(160,140){$u'$}
 \put(210,45){$y$} \put(210,200){$y'$}
 \put(20,50){$v$}  \put(20,205){$v'$}
 \put(75,7){$\a$} \put(33,15){$\b$}
 \put(115,57){$\a$} \put(180,15){$\b$}
 \put(-3,173){$\b$} \put(75,135){$\a$}
 \put(147,173){$\b$} \put(110,205){$\a$}
 \put(-25,75){$\lambda$}\put(155,75){$\lambda$}
 \put(25,115){$\lambda$}\put(205,115){$\lambda$}
\end{picture}
\caption{Elementary cube of the three--dimensional lattice}\label{cube1}
\end{center}
\end{figure}

Elementary building blocks of $\bf D$ are ``cubes''
$(x,u,y,v,x',u',y',v')$, as shown on Fig.\,\ref{cube1}.
This figure is identical with Fig. \ref{cube}, up to notations.
Consider the equation (\ref{basic eq}) on the ``vertical'' face $(x,u,u',x')$:
\[
Q(x,u,x',u';\a,\lambda)=0,
\]
and suppose that it gives $u'$ as a fractional--linear transformation of $x'$:
\[
u'=L(u,x,\a,\lambda)[x'].
\]
Then, due to the three--dimensional consistency, we have:
\[
y'=L(y,u,\b,\lambda)L(u,x,\a,\lambda)[x']=
L(y,v,\a,\lambda)L(v,x,\b,\lambda)[x']\,.
\]
This holds for arbitrary $x'\in\cA$ and for all $\lambda\in\cK$. If now some
structural peculiarities of the matrices $L$ allow us to conclude from the
above that (\ref{zero curv}) holds,
then the matrices $L$ are the transition matrices of a zero curvature
representation. (In the commutative case we could simply normalize the
determinant of $L$ in order to perform the last step.) 

\begin{theorem}
The Hirota equation admits a zero curvature representation with matrices
from the loop group ${\rm GL}(2,\cA)[\lambda]$: the transition matrix along
the (oriented) edge $(x,u)$ carrying the label $\a$ is given by
\begin{equation}
L(u,x,\a;\lambda)=\left(\begin{array}{cc}
\a & -\lambda u \\ \lambda x^{-1} & -\a x^{-1}u\end{array}\right).
\end{equation}
\end{theorem}
{\bf Proof.} Recall that the 11 entries of both matrix products in 
(\ref{zero curv})
are equal to $\a\b-\lambda^2yx^{-1}$. It is easy to see that if for
$\cL,\cM\in{\rm GL}(2,\cA)$ there holds $\cL[\xi]=\cM[\xi]$ for all 
$\xi\in\cA$, and $\cL_{11}=\cM_{11}$, then with necessity $\cL=\cM$. \qed

\subsection{Quantum Hirota equation}

When speaking about solutions of equation like (\ref{basic eq}), one has in
mind a suitably posed initial value problem for it. For a two--dimensional
equation (\ref{basic eq}) Cauchy data (the values of the dependent variable
$x$) should be prescribed along a one--dimensional path, 
i.e. on a sequence of points $\cC=\{\zz_i\}_{i=i_0}^{i=i_1}$,
where $i_0\ge -\infty$, $i_1\le \infty$ and $\zz_i\in V(\cD)$. Whenever such
a path contains three vertices of an elementary quadrilateral from $F(\cD)$,
the equation (\ref{basic eq}) can be applied to get the value of $x$ in the
fourth vertex. This fourth vertex is then said to belong to $\cE(\cC)$,
the {\it evolution set of} $\cC$. (Note that the original three vertices 
are not counted to $\cE(\cC)$.) Continuing this process {\it ad infinitum}, 
we get a full set $\cE(\cC)$ of vertices where the dependent variables 
are defined by a successive application of the equation to the initial 
data along $\cC$. Of course, there are cases when the set $\cE(\cC)$ is empty 
(think, for instance, of the case when $\cC$ is the set of vertices of a 
regular quadratic lattice lying on a coordinate line); one is
interested in $\cC$ with a possibly large $\cE(\cC)$. However,
all data along $\cC$ should be independent. This is formalized in the 
following definition: the path $\cC$ is {\it space--like}, if $\cE(\cC)
\cap\cC=\emptyset$.

\begin{figure}[htbp]
\begin{center}
\setlength{\unitlength}{0.08em}
\begin{picture}(200,100)(0,-20)
  \put(20,0){\circle*{4}} \put(60,0){\circle*{4}}
  \put(100,0){\circle*{4}} \put(140,0){\circle*{4}}
  \put(180,0){\circle*{4}}
  \put(0,20){\circle{4}} \put(40,20){\circle{4}}
  \put(80,20){\circle{4}} \put(120,20){\circle{4}}
  \put(160,20){\circle{4}} \put(200,20){\circle{4}}
  \put(20,40){\circle*{4}} \put(60,40){\circle*{4}}
  \put(100,40){\circle*{4}} \put(140,40){\circle*{4}}
  \put(180,40){\circle*{4}}
  \put(0,60){\circle{4}} \put(40,60){\circle{4}}
  \put(80,60){\circle{4}} \put(120,60){\circle{4}}
  \put(160,60){\circle{4}} \put(200,60){\circle{4}}
  \path(0,20)(20,40) \path(20,40)(40,20)
  \path(40,20)(60,40) \path(60,40)(80,20)
  \path(80,20)(100,40) \path(100,40)(120,20)
  \path(120,20)(140,40) \path(140,40)(160,20)
  \path(160,20)(180,40) \path(180,40)(200,20)
  \path(0,20)(200,20) \path(0,40)(200,40)
\dottedline{2}(0,20)(0,60) \dottedline{2}(200,20)(200,60)
\thicklines
  \path(0,20)(20,0) \path(20,0)(40,20)
  \path(40,20)(60,0) \path(60,0)(80,20)
  \path(80,20)(100,0) \path(100,0)(120,20)
  \path(120,20)(140,0) \path(140,0)(160,20)
  \path(160,20)(180,0) \path(180,0)(200,20)
  \path(0,60)(20,40) \path(20,40)(40,60)
  \path(40,60)(60,40) \path(60,40)(80,60)
  \path(80,60)(100,40) \path(100,40)(120,60)
  \path(120,60)(140,40) \path(140,40)(160,60)
  \path(160,60)(180,40) \path(180,40)(200,60)
  \put(-10,10){$x_0$} \put(15,-10){$x_1$} \put(35,7){$x_2$}
  \put(175,-10){$x_{2N-1}$}
  \put(200,10){$x_{2N}\equiv x_0$}
  \put(-10,67){$\wx_0$}
  \put(15,48){$\wx_1$}
  \put(33,67){$\wx_2$}
  \put(190,67){$\wx_{2N}$}
\end{picture}
\caption{The Cauchy problem on a zigzag} \label{Fig:Cauchy}
\end{center}
\end{figure}
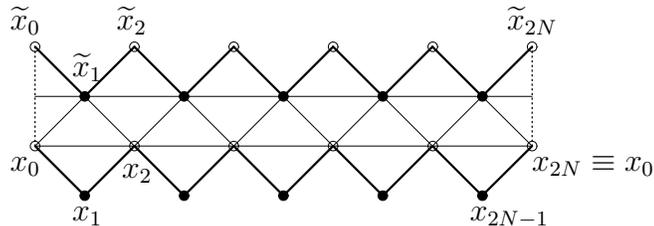

It is well--known that in the case of the regular square lattice  
the zigzag line as in Fig.\,\ref{Fig:Cauchy} is a space--like path
with $\cE$ covering the whole lattice \cite{CNP}, \cite{FV}. Eq. 
(\ref{basic eq}) defines in this case the evolution in the vertical 
direction, i.e. the map $\{x_i\}_{i\in\mathbb Z}\mapsto
\{\widetilde{x}_i\}_{i\in\mathbb Z}$. One often imposes the
periodicity in the horizontal direction with an even period $2N$, in this 
case one is dealing with the regular square lattice on a cylinder rather than 
on the plane.

%
A theory of the quantum Hirota equation on the regular square lattice
was developed in \cite{FV}. As demonstrated there, 
in the non-commutative case the map $\{x_i\}_{i\in\mathbb Z}\mapsto
\{\widetilde{x}_i\}_{i\in\mathbb Z}$ preserves the following Weyl--like
commutation relations:
\begin{equation}\label{Hirota Weyl}
\left\{\begin{array}{cl} [x_i,x_j]=0, & j-i\;\;{\rm even}, \\
x_ix_j=qx_jx_i, & j-i>0 \;\;{\rm odd}.\end{array}\right.
\end{equation}
Actually, this holds for an arbitrary space--like path in an
arbitrary quad--graph $\cD$, and for an arbitrary function $f_{\a\b}$. 
So, this property has, in principle, nothing to do with integrability.
The non--commutative Hirota equation with the Weyl commutation rules
(\ref{Hirota Weyl}) along a space--like path is called the
{\it quantum Hirota equation}. 

We would like to stress that the quantum Hirota equation 
defines an evolution also in the multi--dimensional
situation. This is due to its consistency property proven above for an
arbitrary non--commutative Hirota system. So, in notations of Fig.
\ref{cube}, one can take as a Cauchy path the sequence
$\{x,x_1,x_{12},x_{123}\}$ with the commutation relations
\[
[x,x_{12}]=[x_1,x_{123}]=0,\quad xx_1=qx_1x,\quad xx_{123}=qx_{123}x,\quad
x_{12}x_{123}=qx_{123}x_{12},
\]
and get the values $x_3$ and $x_{23}$ with the commutation relations
\[
[x,x_{23}]=[x_3,x_{123}]=0,\quad xx_3=qx_3x,\quad xx_{123}=qx_{123}x,\quad
x_{23}x_{123}=qx_{123}x_{23},
\]
in two different ways, according to two schemes:
\begin{itemize}
\item $(x_1,x_{12},x_{123})\mapsto x_{13}\,,$ 
      $\;(x,x_1,x_{13})\mapsto x_3\,,$
      $\;(x_3,x_{13},x_{123})\mapsto x_{23}\,.$
\item $(x,x_1,x_{12})\mapsto x_2\,,$ 
      $\;(x_2,x_{12},x_{123})\mapsto x_{23}\,,$
      $\;(x,x_2,x_{23})\mapsto x_3\,.$     
\end{itemize}
Due to the three--dimensional consistency, these two schemes lead to identical
results. This property separates the quantum Hirota equation among all other 
quantum equations of the type (\ref{Hirota type eq}). As demonstrated above,
this property allows us also to {\it derive} the quantum zero curvature 
representation found in \cite{FV}.

It would be desirable to relate this property with another one \cite{FV}, 
namely the Yang--Baxter relation
\[
r(\lambda,u)r(\lambda\mu, v)r(\mu,u)=r(\mu,v)r(\lambda\mu,u)r(\lambda,v)
\]
for the solution of the functional equation 
\[
\frac{r(\lambda,qw)}{r(\lambda,q^{-1}w)}=f_{\lambda}(w),
\]
which also separates the Hirota equation among all other equations
of the type (\ref{Hirota type eq}) with $f_{\a\b}=f_{\b/\a}$.

\section{Non-commutative cross--ratio equation}\label{sect:cr}
\subsection{Equation and its different forms}
Consider the system on a quad--graph $\cD$, consisting of the following
equations on elementary quadrilaterals: 
\begin{equation}\label{cross ratio eq}
 (x-u)(u-y)^{-1}(y-v)(v-x)^{-1}=\frac{\a}{\b},
\end{equation}
where $\a,\b\in\cK$. This equation was considered previously in two particular
settings, when it has important geometrical applications: $\cA=\mathbb C$, 
$\cK=\mathbb C$ (discrete conformal maps; for the case of a regular
square lattice and $\a/\b=-1$ see, e.g., \cite{NC}; for the general case
on arbitrary quad-graphs see \cite{BS}), and $\cA=\mathbb H$, 
$\cK=\mathbb R$ (discrete isothermic surfaces and their Darboux 
transformations, see \cite{BP}, \cite{JHP}). If $\cA=\cC(n)$, the Clifford
algebra over $\cK=\mathbb R$, this equation describes
multi-dimensional isothermic nets, cf. \cite{S}.

In the notations with indices Eq. (\ref{cross ratio eq}) can be put as
\begin{equation}\label{cross ratio eq 2}
\a_1(x_{12}-x_1)(x_1-x)^{-1}=\a_2(x_{12}-x_2)(x_2-x)^{-1}.
\end{equation}
This makes obvious the symmetry of this equation with respect to the
simultaneous flip $x_1\leftrightarrow x_2$, $\a_1\leftrightarrow\a_2$,
as well as to the simultaneous flip $x\leftrightarrow x_{12}$, 
$\a_1\leftrightarrow\a_2$.
Several other forms of this equation and its consequences will be of interest
for us. For instance, we transform (\ref{cross ratio eq 2}) as
\[
\a_1(x_{12}-x)(x_1-x)^{-1}-\a_1=\a_2(x_{12}-x)(x_2-x)^{-1}-\a_2
\]
in order to arrive at the so--called three--leg form:
\begin{equation}\label{cross ratio eq 3leg}
\a_1(x_1-x)^{-1}-\a_2(x_2-x)^{-1}=(\a_1-\a_2)(x_{12}-x)^{-1}.
\end{equation}
Notice that by multiplying this equation by $x_{12}-x$ from the right,
we eventually arrive at
\begin{equation}\label{cross ratio eq r}
\a_1(x_1-x)^{-1}(x_{12}-x_1)=\a_2(x_2-x)^{-1}(x_{12}-x_2),
\end{equation}
which is thus demonstrated to be equivalent to (\ref{cross ratio eq 2}),
the fact non--obvious in the non--commutative context.

One could choose the point $x_{12}$ as the common point of the three legs,
and obtain instead of (\ref{cross ratio eq 3leg}) the equation
\[
\a_2(x_1-x_{12})^{-1}-\a_1(x_2-x_{12})^{-1}=(\a_2-\a_1)(x-x_{12})^{-1}.
\]
Since the right--hand sides of this equation and of Eq. 
(\ref{cross ratio eq 3leg}) coincide, we come to the following
consequence of the basic equation:
\begin{equation}\label{cross ratio eq dual}
\a_1(x_1-x)^{-1}-\a_2(x_2-x)^{-1}=
\a_1(x_{12}-x_2)^{-1}-\a_2(x_{12}-x_1)^{-1}.
\end{equation}

\subsection{Three--dimensional consistency}

\begin{theorem}\label{th cross-rat 3d}
The non--commutative cross--ratio equation is three--dimensionally
consistent.
\end{theorem}
\noindent
{\bf Proof.} We proceed as in the second proof of Theorem 
\ref{Hirota 3-dim compatibility}.
From the cross--ratio equation of the face $(x,x_1,x_{13},x_3)$, 
\[
\a_1(x_{13}-x_1)(x_1-x)^{-1}=\a_3(x_{13}-x_3)(x_3-x)^{-1}
\]
we derive:
\[
\a_1\a_3^{-1}(x_{13}-x_1)(x_1-x)^{-1}(x_3-x)=(x_{13}-x_1)+(x_1-x_3),
\]
which is equivalent to
\[
(x_{13}-x_1)\Big(1+\a_1\a_3^{-1}(x-x_1)^{-1}(x_3-x)\Big)=
x_3-x_1=(x_3-x)+(x-x_1).
\]
This can be put in the matrix form as
\begin{equation}\label{cross ratio aux13}
x_{13}-x_1=L(x_1,x,\a_1,\a_3)[x_3-x],
\end{equation}
where
\begin{equation}\label{cross ratio L prelim}
L(x_1,x,\a_1,\a_3)=\left(\begin{array}{cc} 1 & x-x_1 \\ 
\a_1\a_3^{-1}(x-x_1)^{-1} & 1 \end{array}\right).
\end{equation}
Similarly,
\begin{equation}\label{cross ratio aux23}
x_{23}-x_2=L(x_2,x,\a_2,\a_3)[x_3-x],
\end{equation}
From (\ref{cross ratio aux23}), (\ref{cross ratio aux13}) we derive, by the 
shift in the direction of the first, resp. the second coordinate axis, 
the expressions for $x_{123}$ obtained by
the first, resp. the second scheme above:
\begin{eqnarray}
x_{123}-x_{12} & = & L(x_{12},x_1,\a_2,\a_3)[x_{13}-x_1],
\label{cross ratio aux123'}\\
x_{123}-x_{12} & = & L(x_{12},x_2,\a_1,\a_3)[x_{23}-x_2].
\label{cross ratio aux123''}
\end{eqnarray}
Substituting (\ref{cross ratio aux13}), (\ref{cross ratio aux23}) on the
right--hand sides of (\ref{cross ratio aux123'}), (\ref{cross ratio aux123''}),
respectively, we represent the equality we want to demonstrate in the 
following form:
\begin{eqnarray}\label{cross rat to prove}
\lefteqn{L(x_{12},x_1,\a_2,\a_3)L(x_1,x,\a_1,\a_3)[x_3-x]}\nonumber\\
& = & L(x_{12},x_2,\a_1,\a_3)L(x_2,x,\a_2,\a_3)[x_3-x].
\end{eqnarray}
This is a consequence of a stronger claim:
\begin{equation}\label{cross rat to prove 1}
L(x_{12},x_1,\a_2,\a_3)L(x_1,x,\a_1,\a_3)=L(x_{12},x_2,\a_1,\a_3)
L(x_2,x,\a_2,\a_3).
\end{equation}
Indeed, the $12$ entries on the both sides are equal to $x-x_{12}$.
Equating the $11$ entries is equivalent to Eq. (\ref{cross ratio eq 2}),
equating the $22$ entries is equivalent to the (inverted) Eq.
(\ref{cross ratio eq r}), and equating the $21$ entries is equivalent
to Eq. (\ref{cross ratio eq dual}). This finishes the proof. \qed
\medskip

{\bf Remark.} As in the case of the Hirota equation, this proof does not
lead to an expression for $x_{123}$ which would make the claim self--evident.
However, also in this case the three--leg form of equations comes to help.
Namely, summing up the equations
\[
\a_i(x_{123}-x_{jk})^{-1}-\a_j(x_{123}-x_{ik})^{-1}=
(\a_i-\a_j)(x_{123}-x_k)^{-1},
\]
we come to the equation
\begin{equation}\label{cross ratio x123 sym}
(\a_2-\a_3)(x_{123}-x_1)^{-1}+(\a_3-\a_1)(x_{123}-x_2)^{-1}
+(\a_1-\a_2)(x_{123}-x_3)^{-1}=0,
\end{equation}
which makes two things obvious: first, that $x_{123}$ depends only on
$x_1$, $x_2$, $x_3$ and not on $x$ (tetrahedron property), and second,
the symmetry of the resulting $x_{123}$ with respect to permutations of 
indices $(1,2,3)$. Comparing (\ref{cross ratio x123 sym}) with 
(\ref{cross ratio eq 3leg}), we see that the former equation is again
of the cross--ratio type: it is equivalent, e.g., to
\begin{equation}\label{cross ratio x123 cross rat} 
(x_{123}-x_1)(x_1-x_3)^{-1}(x_3-x_2)(x_2-x_{123})^{-1}=
(\a_2-\a_3)(\a_3-\a_1)^{-1}.
\end{equation}

\subsection{Zero curvature representation}

Setting $\a_3=\lambda^{-1}$ in the proof of Theorem \ref{th cross-rat 3d},
we come to the following statement.
\begin{theorem}
The cross--ratio equation admits a zero curvature representation with 
matrices from the loop group ${\rm GL}(2,\cA)[\lambda]$: the transition 
matrix along the (oriented) edge $(x,u)$ carrying the label $\a$ is given by
\begin{equation}\label{cross ratio L}
L(u,x,\a;\lambda)=
\left(\begin{array}{cc}
1 & x-u \\ \lambda\a (x-u)^{-1} & 1\end{array}\right).
\end{equation}
\end{theorem}
An essential point is, we stress it again, that this zero curvature 
representation was {\it derived} from the equations of the system, 
without any additional information. Moreover, the proof of Theorem
\ref{th cross-rat 3d} shows again that the the zero curvature representation
not only follows from the three--dimensional consistency, but is, in turn,
instrumental in establishing it.

\section{Concluding remarks}
 
 The present paper has to be considered in the context of the ongoing
 study of non-commutative integrable systems \cite{EGR1}, \cite{EGR2},
 \cite{K} which puts quantum integrable systems on a more
 general basis (see also \cite{KL}, \cite{GR}). We expect that discrete
 integrable non-commutative systems of the sort considered in this paper 
 are of a fundamental importance, just like it is the case in the commutative
 context \cite{BS}. It will be important to extend the classification
 results of \cite{ABS} to the non-commutative case, and to get complete 
 lists of discrete integrable equations. Also, a more thorough
 understanding of the quantum case and the origin of its specific 
 (Yang--Baxter) structures is desirable. This could lead also to a new
 approach to classification of solutions of the Yang--Baxter equation.

\paragraph{Acknowledgment.} This research was partly supported by
DFG (Deutsche Forschungsgemeinschaft) in the frame of SFB 288
''Differential Geometry and Quantum Physics''.



\begin{thebibliography}{99}
\setlength{\itemsep}{0mm}
\bibitem{ABS} V.E.Adler, A.I.Bobenko, Yu.B.Suris. Classification of 
 integrable equations on quad--graphs. The consistency approach.
 {\em Preprint}, {\tt http://www.arXiv.org/abs/nlin.SI/0202024}.

\bibitem{BP} A.I.Bobenko, U.Pinkall. Discrete isothermic surfaces. 
{\it J. Reine Angew. Math}, 1996, {\bf 475}, 187--208.

\bibitem{BS} A.I.Bobenko, Yu.B.Suris. Integrable systems on quad-graphs.
 {\em Int. Math. Res. Notices}, 2002, No 11, 573--611.

\bibitem{CNP} H.W.Capel, F.W.Nijhoff, V.G.Papageorgiou.
 Complete integrability of Lagrangian mappings and lattices of KdV type.
 {\em Phys. Lett. A},  1991, {\bf 155}, 377--387.
 
\bibitem{EGR1} P.Etingof, I.Gelfand, V.Retakh. Factorization of 
 differential operators, quasideterminants, and nonabelian Toda field 
 equations. {\it Math. Res. Lett.}, 1997, {\bf 4}, 413--425.  

\bibitem{EGR2} P.Etingof, I.Gelfand, V.Retakh. Nonabelian integrable systems, 
 quasideterminants, and Marchenko lemma. {\it Math. Res. Lett.}, 1998, {\bf 5},
 1--12.
 
\bibitem{FKV} L.D.Faddeev, R.M.Kashaev, A.Yu.Volkov. Strongly coupled
 quantum discrete Liouville theory. I: Algebraic approach and duality.
 {\it Commun. Math. Phys.}, 2001, {\bf 219}, 199--219.  

\bibitem{FV} L.D.Faddeev, A.Yu.Volkov. Hirota equation as an example of
 an integrable symplectic map. {\em Lett. Math. Phys.}, 1994, {\bf 32},
 125--135.
 
\bibitem{GR} I.Gelfand, V.Retakh. Quasideterminants. I. {\it Selecta Math.}, 
 1997, {\bf 3}, 517--546. 

\bibitem{H} R.Hirota. Nonlinear partial difference equations. I. A
 difference analog of the Korteweg--de Vries equation.
 III. Discrete sine-Gordon equation. {\em J. Phys. Soc. Japan}, 1977,
 {\bf 43}, 1423--1433, 2079--2086.
 
\bibitem{JHP} U.Hertrich-Jeromin, T.Hoffmann, U.Pinkall. A discrete 
 version of the Darboux transform for isothermic surfaces. -- In:
 {\it Discrete Integrable Geometry and Physics}, Ed. A.I.Bobenko,
 R. Seiler, Oxford: Clarendon Press, 1999, 59--79.
 
\bibitem{KL} D.Krob, B.Leclerc. Minor identities for quasi-determinants 
 and quantum determinants. {\it Comm. Math. Phys.}, 1995, {\bf 169}, 
 1--23. 
 
\bibitem{K}  B.A.Kupershmidt. KP or mKP. Noncommutative mathematics of 
 Lagrangian, Hamiltonian, and integrable systems. Providence, RI: AMS, 2000.

\bibitem{N} F.W.Nijhoff. Lax pair for the Adler (lattice Krichever-Novikov)
 system. {\it Phys. Lett. A}, 2002, {\bf 297}, 49--58.

\bibitem{NC} F.W.Nijhoff, H.W.Capel. The discrete Korteweg-de Vries equation.
 {\em Acta Appl. Math.}, 1995, { \bf 39}, 133--158.

\bibitem{S}  W.K.Schief. Isothermic surfaces in spaces of arbitrary dimension: 
 integrability, discretization, and B\"acklund transformations -- a discrete 
 Calapso equation. {\it Stud. Appl. Math.}, 2001, {\bf 106}, 85--137.



\end{thebibliography}
\end{document}